\documentstyle[11pt]{article}
\newcommand{\be}{\begin{equation}}
\newcommand{\ee}{\end{equation}}
\newcommand{\ba}{\begin{eqnarray}}
\newcommand{\ea}{\end{eqnarray}}
\newcommand{\nn}{\nonumber}

\begin{document}

\begin{flushright}
\vspace*{-1.5cm}
        IEKP-KA/94-08    \\
         hep-ph/9406419  \\
        June, 1994        \\
\end{flushright}
\vspace{2.1cm}
\begin{center}

{\bf Bounds on the Ratio of Higgs Vacuum Expectation Values  \\
     in the Minimal Supersymmetric Standard  Model \\
     from the Top and Bottom Quark Masses.}

\bigskip

{V.A. Bednyakov$^*$\footnotemark[1],
W. de Boer\footnotemark[2], S.G. Kovalenko$^*$
\footnotemark[3]}
\bigskip

{\it Inst. f\"{u}r Experimentelle Kernphysik, Univ. of Karlsruhe\\
Postfach 6980, D-76128 Karlsruhe, Germany}

$^*${\it Joint Institute for Nuclear Research, Dubna, Russia}

\end{center}
\vspace*{2cm}

\begin{abstract}
The top
quark mass  values from the  CDF Collaboration  and the
  precision electroweak data at LEP
combined with the bottom mass value
allow to establish  bounds on the
parameter $\tan\,  \beta$ in the minimal supersymmetric standard model
 (MSSM)
independent of the soft supersymmetry breaking parameters.
We find:
 $0.96 \leq \tan\,  \beta \leq  52.2~{\rm at~the ~ 95\%~C.L.}$, which is
mostly
the region $\tan\, \beta >1$
where   radiative electroweak symmetry
breaking is possible.
Assuming electroweak symmetry breaking
leads to: $\tan\, \beta=1.2\pm 0.2$, if one neglects the high
  $\tan\, \beta$ solution, which is unlikely due to proton lifetime
limits.
\end{abstract}

\bigskip

\footnotetext[1]{e-mail: BEDNY@NUSUN.JINR.DUBNA.SU}
\footnotetext[2]{BITNET: DEBOERW@CERNVM}
\footnotetext[3]{e-mail: KOVALEN@NUSUN.JINR.DUBNA.SU}

\newpage

\section{Introduction}

The minimal supersymmetric standard model (MSSM) \cite{susyrev}
requires
a minimum of two Higgs doublets: one doublet $H_2$ coupling to the up-type
quarks and one doublet $H_1$ coupling to the down-type quarks and leptons.
After electroweak symmetry breaking the neutral components of both Higgs
doublets develop a non-zero vacuum expectation value (v.e.v.); the ratio
of these v.e.v.'s is denoted by $\tan\,  \beta = <H_2^0>/<H_1^0>$.
The interaction with $<H_2^0>$ gives masses to the up-type quarks
while the one with $<H_1^0>$ gives masses to the down-type quarks and
to charged leptons.

In this note we point out that recent measurements of the top quark mass
\cite{CDF}, \cite{LEP} establish a lower bound on $\tan\,  \beta$
independent of
a specific scenario or supergravity inspired soft breaking terms.

Only two assumptions are necessary for this purpose. The first is
the absence of new physics, except supersymmetry, between the Fermi
scale $Q \approx M_Z$ and the unification scale $Q \approx
 M_X \approx 10^{16}$GeV.
The second assumes the theory to be perturbative up to this unification
scale. Both assumptions are always implied in the MSSM.

{}From the fact that only supersymmetry governs physics up to $M_X$
allows one to calculate the coupling constants at any point from the
known renormalization group equations (RGE). Extrapolating the coupling
constants from $M_Z$ to $M_X$ and requiring unification allows one
to estimate the SUSY breaking scale $M_{susy}$, the unification point
$M_X$ and the common gauge coupling at this point $\alpha_{GUT}$
\cite{de_boer}:

\be \label{Unif}
 \alpha_1(M_X) = \alpha_2(M_X) =  \alpha_3(M_X) = \alpha_{GUT}
\ee

with

\ba \label{Sol1}
M_X& =&  10^{16.2 \pm 0.3 \pm 0.1}~ \mbox{GeV}\\
\alpha_{GUT}^{-1}& =& 24.5_{-0.7}^{+1.4} \label{Sol2}\\
 M_{susy}& =& 10^{2.6 \pm 0.9 \pm 0.4} \mbox{GeV}\label{Sol3},

\ea
where $\alpha_i = C_i g_i^2/4\pi$; $C_1 = 5/3, C_2 = C_3 = 1$ and
$g_{3,2,1}$  are $SU_3\times SU_2 \times U_1$ gauge coupling constants.

Thus in the MSSM automatically a new mass scale $M_X$ appears where
unification of the standard gauge group to the large one may happen.
Accepting this point, we consider the 1-loop RGE for the top and bottom
Yukawa coupling constants and use their general properties to obtain
bounds on these coupling constants from the known top and bottom
quark masses.

\section{Infrared Fixed Points in the Renormalization Group
Equations}

The one loop RGE for the Yukawa couplings of the top and bottom
quarks have the
following form \cite{RGE}:
\ba \label{Ytb}
\frac{dY_t}{dt} & = & Y_{t}(t) G_{t}(t) - 6 Y_t^2 - Y_t Y_b,\\
\frac{dY_b}{dt} & = & Y_{b}(t) G_{b}(t) - 6 Y_b^2 - Y_t Y_b.
\label{Ytb1}
\ea
Here $t = \ln (\frac{M_X^2}{Q^2})$ and
\ba \label{G_q}
&&G_{q}(t) = \sum_{i=1}^{3} k_{i}^{(q)}  \tilde{\alpha}_i,\ \ \ \
\ \ \ \mbox{q = t, b} \\
&&k_1^{(t)} = \frac{13}{15}, \ \ \ k_1^{(b)} = \frac{7}{15},\\
&&k_2^{(t)} = k_2^{(b)} = 3, \ \ \ k_3^{(t)} = k_3^{(b)} = \frac{16}{3}.
\ea
with $\tilde{\alpha}_i = \alpha_{i}/4\pi$.
The scale dependence of the gauge coupling constants at the 1-loop level is
given by
\be \label{alph}
\tilde{\alpha}_i(t)  = \frac{\tilde{\alpha}_{GUT}}
	                 {1 + \tilde{\alpha}_{GUT}\ b_i\ t}

\ee
The RGE coefficients in the MSSM are $b_1 = 33/5, b_2 = 1, b_3 = -3$.

{}From eqns. \ref{Ytb}-\ref{Ytb1}  the following inequalities

can be easily obtained \cite{Bagger}:
\ba \label{lim}
\frac{dY_t}{dt} & \leq & Y_{t}(t) G_{t}(t) - 6 Y_t^2,\\
\frac{dY_b}{dt} & \leq & Y_{b}(t) G_{b}(t) - 6 Y_b^2. \label{lim1}
\ea
These eqns. give upper bounds on $Y_t$ and $Y_b$ at some fixed scale $Q$:
\ba \label{boundY}
Y_t(t) \leq \tilde{Y}_t(t) &=& \frac{Y_t(0) E_t(t)}{1 + Y_t(0) 6 F_t(t)}\\
Y_b(t) \leq \tilde{Y}_b(t) &=& \frac{Y_b(0) E_b(t)}{1 + Y_b(0) 6 F_b(t)}.
\label{boundY1}
\ea
Here $Y_{t,b}(0)$ are the
initial values of the top and bottom Yukawa
coupling constants at $Q = M_X$ and the functions $E$ and $F$
are defined as:
\ba \label{e_f}
E_q(t) &=& \prod_{i=1}^{3} (1 + \beta_i t)^{k_i^{(q)}/b_i}\\
F_q(t) &=& \int_{0}^{t} d t' E_q(t').
\ea
Here  $\beta_i = b_i~\tilde{\alpha}_{GUT}$.

The limiting functions $\tilde{Y}_{t,b}(t)$ have the well known
infrared fixed points $Y_{t,b}^{fix}(t)$ in the limit of large
initial values $Y_{t,b}(0)$. In this  limit one obtains
from eqns. \ref{boundY}-\ref{boundY1} upper bounds at a fixed scale
$Q$:
\ba \label{FIX}
\tilde{Y}_t \leq Y_t^{fix} &=& \frac{E_t(t)}{6 F_t(t)}\\
\tilde{Y}_b \leq Y_b^{fix} &=& \frac{E_b(t)}{6 F_b(t)}.
\label{FIX1}
\ea
%Combining eqs. \ref{boundY} and \ref{FIX} one can obtain the absolute
%upper bounds for the top and bottom Yukawa coupling constants at
%the Fermi scale $Q = M_Z$.
With the values of $M_X$ and $\alpha_{GUT}$ obtained from
the gauge coupling constants unification (eqns. \ref{Sol1}-\ref{Sol2}),

we obtain the following limits:
\ba \label{Fermi}
Y_t \leq Y_t^{fix}  \approx 8\cdot 10^{-3}\\
Y_b \leq Y_b^{fix}  \approx 8.4\cdot 10^{-3}.\label{Fermi1}
\ea
Now we are in a position to obtain upper and lower bounds for
$\tan\, \beta$.

\section{Bounds on  $\tan\, \beta$ from the IR fixed points}

The top and bottom Yukawa couplings $Y_t$ and $Y_b$ are related
directly to the running top and bottom quark masses:
\ba \label{y_m}
 m_{t}^2(t=m_{t}^2) &=& (4\pi)^2 Y_{t}(t = m_t^2)\ v^2 \ \sin^2 \beta   \\
 m_{b}^2(t=m_{b}^2) &=& (4\pi)^2 Y_{b}(t = m_b^2)\ v^2 \ \cos^2 \beta.
\ea
Here $M_Z^2 = v^2 g_2^2/ (2 cos^2\theta_W)$.
Replacing the Yukawa couplings by the upper limits $Y_i^{fix}$
(eqns. \ref{Fermi}-\ref{Fermi1})
one obtains the following inequalities:
\be \label{yy_mt}
m_{t}(t=m_{t}^2)  \leq  198~\mbox{GeV} \, \sin\, \beta
\ee
\be \label{yy_mb}
m_{b}(t=m_{b}^2)  \leq  207~\mbox{GeV} \, \cos\, \beta.
\ee

The running quark masses $m_{t,b}$ are related in the
$\overline{MS}$ renormalization scheme
to the physical "pole" masses $M_{t,b}$ by \cite{Gasser}:
\be\label{pole}
m_i(t = m_{i}^2) = M_i \left(1 - \frac{4}{3} \frac{\alpha_s}{\pi} -
	12.4 \left(\frac{\alpha_s}{\pi}\right)^2\right)

\ee
where the QCD  running coupling constant $\alpha_{s}$ is taken
at the scale $t = m_i^2$.
For the top bottom quark masses one finds:
\ba\label{pole_num}
m_t \approx 0.935 ~M_t,\\ \nn
m_b \approx 0.844 ~M_b.
\ea
{}From eqns. \ref{yy_mt}-\ref{pole_num} we obtain
 for the $\tan\, \beta$ parameter
\be \label{Const}
\frac{M_t}{\sqrt{(212\ \mbox{GeV})^2 - M_t^2}}
\leq \tan\, \beta \leq \frac{\sqrt{(245\ \mbox{GeV})^2 - M_b^2}}{M_b}.
\ee

Combining the results on the top mass from the CDF collaboration
($M_t = 174\pm 10 {}_{-12}^{+13} \mbox{GeV}$ \cite{CDF}) and
the electroweak data from LEP and SLC
($M_t = 177\pm 11 {}_{-19}^{+18} \mbox{GeV}$ \cite{LEP}) one obtains
$M_t=175\pm 14$ GeV. The central value from LEP and SLC is quoted for
a Higgs mass of 300 GeV and the last errors correspond to
a Higgs mass variation between 100 and 1000 GeV.
Ellis, Fogli and Lisi \cite{ellis} leave the Higgs mass free in their
analysis.  They perform
the fit in the SM and MSSM,
 including  data from LEP, SLC, and the Tevatron, as well as lower energy data.
  Their fit prefers a Higgs
 mass of the order of 100 GeV in agreement with  the expectations of the
 MSSM. A low Higgs mass corresponds to a lower top mass. They find:
\be \label{CDF}
M_t = 162\pm 9\, \mbox{GeV}
\ee
Since we work  in the framework of the MSSM, we will use this value.
Inserting the 95\% C.L. lower limits on the masses $M_t = 147$ GeV and
$M_b = 4.7$ GeV \cite{Part} in eq. \ref{Const} yields:
\ba \label{Tan_CDF}
      0.96\ \leq \tan\, \beta  \leq \  52.1 \ \ \mbox{ at 95\% C.L.}
\ea
It is worthwhile noticing that this
%%%% modification. %%%%%%%%%%%%%%%%%%%%
practically  excludes

%includes
 the region $\tan\, \beta < 1$ where   radiative electroweak symmetry
breaking  due to radiative
corrections from the large top Yukawa coupling
is impossible \cite{ewbr,susyrev}.
%%%%%%%%%%%%%%%%%%%%%%%%%%%%%%%%%%%%%%%
In this respect one can regard the heavy
top quark mass as an indirect hint  in favour of this scenario of
mass generation.

Requiring electroweak symmetry breaking implies a large value
of  the top Yukawa coupling at the GUT scale,
in practice   close to $Y_t^{fix}$
\cite{wdb1}.
In this case eq. \ref{yy_mt}  should have an equal sign
and the value of $\tan\, \beta$
can be determined immediately from the top mass,
 if one neglects a high $\tan\, \beta$
solution \cite{lanpol},
 which is unlikely because
of a too fast proton decay \cite{nath,wdb1}.
%the contribution from the bottom
%Yukawa coupling to the Higgs potential.
For $M_t=162\pm 9$ GeV one finds:
\be  \tan\, \beta =1.2\pm 0.2 \ee
 The direct measurement from CDF alone ($M_t=174\pm17$ GeV)
corresponds to \mbox{\tan \beta=1.4^{+0.7}_{-0.3}$}.

 \bigskip

\centerline{\bf ACKNOWLEDGMENTS}

We thank R. Ehret, D. Kazakov  and W. Oberschulte-Beckmann
 for   helpful discussions.
The research described in this publication was made possible in part by Grant
No. RFM000 from the International Science Foundation, by support from
  the Human Capital and Mobility Program
 (Contract ERBCHRXCT 930345)  from the European Communities, and by
support from the German Bundesministerium f\"ur Forschung und Technologie
(BMFT)
(Contract 05-6KA16P).1.
 \bigskip\\


\begin{thebibliography}{99}
\bibitem{susyrev}
{\rm Reviews and original references can be found in:\\ H.E. Haber, Lectures
  given at Theoretical Advanced Study Institute, University of Colorado, June
  1992, Preprint Univ. of Sante Cruz, SCIPP 92/33; see also SCIPP 93/22;\\ {\it
  Perspectives on Higgs Physics}, G. Kane (Ed.), World Scientific, Singapore
  (1993);\\ {\it Int. Workshop on Supersymmetry and Unification}, P. Nath
  (Ed.), World Scientific, Singapore (1993);\\ {\it Phenomenological Aspects of
  Supersymmetry}, W. Hollik, R. R\"uckl and J. Wess (Eds.), Springer Verlag
  (1993);\\ R. Barbieri, Riv. Nuovo Cim. {\bf 11} (1988) 1;\\ A.B. Lahanus and
  D.V. Nanopoulos, Phys. Rep. {\bf 145} (1987) 1;\\ H.E. Haber and G.L. Kane,
  Phys. Rep. {\bf 117} (1985) 75;\\
P. Nath, R. Arnowitt, and A.H. Chamseddine,  \\Applied  N=1 Supergravity,
World Scientific, Singapore, 1984;\\
M.F. Sohnius, Phys. Rep. {\bf 128} (1985)
  39;\\ H.P. Nilles, Phys. Rep. {\bf 110} (1984) 1;\\
P. Fayet and S. Ferrara,
  Phys. Rep. {\bf 32} (1977) 249.
}
 \bibitem{CDF} CDF collaboration, F.Abe, et al., FERMILAB-PUB-94/116-E,
May, 1994; Nature, 368 (1994) 783; 805.
\bibitem{LEP} The LEP Collaboration  ALEPH, DELPHI, L3 and OPAL and the
SLC Collaboration SLD; preliminary data
presented at the Moriond Conference by the
LEP electroweak Working Group: Internal Note LEPEWWG/94-01.
\bibitem{ellis} J. Ellis, G.L. Fogli and E. Lisi,
CERN Preprint CERN-TH 7261/94.
\bibitem{de_boer} U. Amaldi, W. de Boer, and H. F\"{u}rstenau,
Phys.Lett. B260 (1991) 447;
The values have been updated using
newer values for the coupling constants,
see W. de Boer, Grand Unified Theories and Supersymmetry in Particle
Physics and Cosmology, IEKP-KA/94-01,
Progr. in Nucl. and Particle Phys. {\bf 33} (1994) 201.

\bibitem{RGE}  L.E. Ib\'a\~{n}ez, C. Lop\'ez, Phys.Lett. B126 (1983) 54;\\
Nucl. Phys. B233 (1984) 511;\\
L.E. Ib\'a\~{n}ez, C. Lop\'ez and  C.Mu\~{n}oz, Nucl. Phys. B256 (1985)
218.
\bibitem{Bagger} J. Bagger, S. Dimopoulos and E. Masso,    Phys. Lett.
B156 (1985) 357;  Phys. Rev. Lett. 55 (1985) 920.
\bibitem{Gasser} J. Gasser and H. Leutwyler, Phys. Rep. C87 (1982) 77;\\
S. Narison, Phys.Lett. B216 (1989) 191; \\
N. Gray, D.J. Broadhurst, W. Grafe and K. Schilcher, Z. Phys. C48 (1990) 673.
%22
\bibitem{Part} Review of Particle Properties, Phys. Rev. D45, No.11 (1992).
\bibitem{ewbr}
{\rm K. Inoue, A. Kakuto, H. Komatsu, and S. Takeshita, Prog. Theor. Phys. {\bf
  68} (1982) 927; ERR. ibid. {\bf 70} (1983) 330;\\ L.E. Ib\'a\~nez, C.
  Lop\'ez, Phys. Lett. {\bf 126B} (1983)\\ 54; Nucl. Phys. {\bf B233} (1984)
  511;\\ L. Alvarez-Gaum\'e, J. Polchinsky, and M. Wise, \\Nucl. Phys. {\bf
221}
  (1983) 495;\\ J. Ellis, J.S. Hagelin, D.V. Nanopoulos, K. Tamvakis, \\Phys.
  Lett. {\bf 125B} (1983) 275;\\ G. Gamberini, G. Ridolfi and F. Zwirner, Nucl.
  Phys. {\bf B331} (1990) 331, \\
   R. Arnowitt and P. Nath, Phys. Rev. {\bf D46} (1992) 3981.}
\bibitem{wdb1}   W. de Boer, R. Ehret,
W. Oberschulte-Beckmann, D. Kazakov,
  IEKP-KA/94-05, hep-ph/9405342.
  \bibitem{lanpol}
  P. Langacker and N. Polonsky, Univ. of Penssylvania Preprint UPR-0594T,
  Feb. 1994.
\bibitem{nath}
R.~Arnowitt and P.~Nath.
\newblock {\rm Phys.~Rev.~Lett.~{\bf 69} (1992) 725; Phys.~Lett.~{\bf B287}
  (1992) 89; Phys.~Lett.~{\bf B289} (1992) 368; Phys.~Lett.~{\bf B299} (1993)
  58, ERRATUM-ibid.~{\bf B307} (1993) 403; Phys.~Rev.~Lett.~{\bf 70} (1993)
  3696; \\CTP-TAMU-23/93 (1993)}
\newblock and references therein.


\end{thebibliography}
\end{document}